 \documentclass{emulateapj}
 \usepackage{apjfonts}

\usepackage{graphicx,epsfig,natbib}
\shorttitle{A GRBR in our Galaxy: HESS J1303-631}
\shortauthors{Atoyan, Buckley \& Krawczynski }

\received{ 2005 December 21; accepted 2006 March 30}

\begin{document}

\title{A gamma-ray burst remnant in our Galaxy: HESS J1303-631}

\author{Armen Atoyan$^1$, James Buckley$^2$ and Henric Krawczynski$^2$}

 \affil{$^1\,$CRM, Universit\'e de Montr\'eal, C.P. 6128, 
 Montr\'eal, Canada H3C 3J7; atoyan@crm.umontreal.ca, }
\affil{$^2\,$Washington University, Physics Department, CB 1105,  
 St. Louis, MO 63130, USA;  \footnotesize
  buckley@wuphys.wustl.edu, krawcz@physics.wustl.edu }

\begin{abstract}
We present an investigation of the multiwavelength data on 
HESS~J1303-631, an unidentified TeV source serendipitously discovered in the 
Galactic plane by the HESS collaboration. Our results strongly
suggest the identification of this particular source as the remnant of a 
Gamma-Ray Burst (GRB) that happened some few tens of thousands years 
ago in our Galaxy at a distance on the order of $\gtrsim \! 10\,$kpc from us.
We show through detailed calculations of particle diffusion, interaction 
and radiation processes of relativistic particles in the interstellar medium,
that it is possible for a GRB remnant (GRBR) to
be a strong TeV emitter with no observable synchrotron emission.  
We predict spectral and spatial
signatures that would unambiguously distinguish GRBRs from ordinary supernova
remnants, including: (1) large energy budgets inferred from their TeV 
emission, but at the same time (2)
suppressed fluxes in the radio through GeV wavebands; (3) extended
center-filled emission with an energy-dependent 
spatial profile; and (4) a possible elongation in the direction 
of the past pair of GRB jets. While GRBRs can best be detected by
ground-based gamma-ray detectors, the future GLAST 
mission will play a crucial role in confirming the predicted 
low level of GeV emission.  
\end{abstract}


\maketitle

\section{Introduction and Outline}

So far, GRBs have only been identified at cosmological distances.  
It is believed that GRBs are caused by highly relativistic outflows
with bulk Lorentz factors $\Gamma \gtrsim 100$ that form a pair of
opposite jets. 
Even after correcting for narrow beaming, 
the energy radiated by long-duration ($\geq 2\,\rm s$) GRBs
typically is $\sim \! 10^{51}\,\rm erg$,
and the estimated kinetic energy of the jets reaches values 
of $\sim \! 10^{52}\,\rm erg$ \citep{mesz02,bfk03,bkf04}. 
Relativistic shocks  convert   
(accelerate) 
{\it all} particles from the incoming plasma they encounter 
to relativistic energies by randomizing their velocities in the 
comoving frame. 
In the stationary frame the mean energy per particle is 
$\overline{E} \simeq m_pc^2 \Gamma^2/2$ \citep{bmk76}.
The total energy of cosmic rays (CRs) produced by a GRB can thus 
be up to $\sim \! 100$ times higher 
than the  $\sim\! 10^{50}\,\rm erg$ produced by nonrelativistic shocks 
of typical supernova remnants (SNR) \citep{ber90}. 
GRBs are prime candidate sources for the extragalactic component of CRs, 
and they may also be sources of ultrahigh energy CRs 
in our Galaxy \citep{der02}. Calculations taking into account
the star formation history and the GRB beaming factor 
predict a Galactic rate of one long 
 GRB every 3000 to $10^5$ years \citep{wda04,dh05}.
A GRB that occurred $\sim \! 1\,\rm Myr$ ago at $\lesssim \! 1\,$kpc from
us could explain  
the knee in the observed 
CR spectrum at $E\gtrsim 10^{15}\,\rm eV$ \citep{wda04}. 
Ohers have suggested that giant radio shell objects 
in the Milky Way, such as the  
SNR W49B \citep{ikm04} or 
kpc-size HI super-shells \citep{efr98} could be remnants of 
GRBs.

The HESS 
(High Energy Stereoscopic System) collaboration has recently discovered a 
population of $\gamma$-ray sources in the Galactic 
plane \citep{science,Ginner} that are TeV-bright 
but are often quiet in all other wavelengths and remain unidentified.
All previously detected TeV sources show strong 
X-ray emission which is the synchrotron counterpart of the 
Inverse Compton (IC) TeV radiation from multi-TeV electrons. 
The suppression of low-frequency flux is unexpected within conventional 
models. \citet{mitra} has proposed that this feature points to 
radiation production in the deep gravitational potential wells of black holes 
formed by Galactic GRBs. However, the extended appearance of these sources 
seems to contradict this explanation. 

The estimated rate of Galactic GRBs implies a 
likelihood of between one and several ``young'' (age $t\! \gtrsim 10^4\,$yr), 
GRB remnants in the interstellar medium (ISM) at distances 
up to $d \! \sim \! 10$-20\,kpc  
from us. Some $\gtrsim 10^4$ years after their acceleration 
by the relativistic shocks of GRBs, multi-TeV particles would 
diffuse over distances of about $\gtrsim \! 100\,$pc. Therefore, if detected 
in TeV $\gamma$-rays, Galactic GRBRs should appear as large 
plerionic (`center-filled') nebulae in the ISM.  

While probably most of the unidentified TeV sources would
eventually be associated with conventional Galactic objects,  
here we argue that at least one source, HESS~J1303-631 \citep{hess1303}, 
shows features that strongly suggest its identification with the remnant 
of a GRB that happened  $\gtrsim 10^4 \,(+ \! d/c)$ years ago
in our Galaxy  at a distance $d\! \simeq (10$-15)\,kpc from us. 
We show that this can
explain both the total CR energy in the source, and its non-detection 
below TeV energies. 
Furthermore, we discuss observational signatures that 
can be  used to identify GRBRs unambiguously.

\section{Investigation of the TeV energy spectrum} 
 
Among the unidentified TeV sources discovered by HESS, 
J1303-631  is the brightest source  with
the most detailed data \citep{hess1303}. 
The measured $(0.38$-$12)$\,TeV differential spectrum, approximated as  
a power-law $F(E) \propto E^{-\alpha_\gamma}$, has  
a mean spectral index $\alpha_{\gamma} \approx 2.44$ and an 
integrated energy flux 
$f_{E} \! \approx \! 1.87 \!\times \! 10^{-11}\,\rm erg\, cm^{-2}\, s^{-1}$. 
So far, no emission has been detected in any other wave band,
including the X-ray band \citep{mh05}.  The source is extended and
is approximated, within current statistical errors, by a spherically 
symmetric 2-dimensional Gaussian distribution with inherent width  
$\theta_{\rm src} = 9^{\prime}.6$ \citep{hess1303}.
The center-filled appearance of the source and the gradual steepening 
of the TeV spectrum suggest the emission origin   
from a population of high-energy particles injected into the 
ISM from a point source and evolved by energy-dependent diffusion. 

Absence of a detectable synchrotron flux implicitly suggests a hadronic 
origin of the TeV radiation.  
Since the two $\gamma$-rays from a $\pi^0$-decay each carry a fraction 
of the incident proton energy of about 
$E_\gamma \! \simeq \! 0.075 \,E_p$, 
the observed flux requires  
$U_{p}\simeq 4\pi d^2 f_{E} t_{pp}\eta_{0}^{-1}$ total energy
in protons with $E_{p}\! \gtrsim 5 \,\rm TeV$.
Here, $t_{pp}=(K_{pp}\sigma_{pp} c n_{\rm H})^{-1}$ is the $pp$-cooling time, 
and $\eta_0 \simeq 1/3$ is the fraction of the lost proton   
energy that is converted into $\pi^0$-mesons. 
Substitution of the cross-section $\sigma_{pp}\simeq 40\,\rm mb$ and 
the inelasticity $K_{pp}\simeq 0.5$ gives
\begin{equation}
U_{p}\simeq 1.2 \times 10^{49}d_{\rm kpc}^2 n_{\rm H}^{-1} (1+S)\;
\rm erg \,.
\end{equation}
Here $d_{kpc} \! = \! d/1\,\rm kpc$, and $S$ accounts for 
protons with $E\! \leq \! 5\,$TeV.
For standard SNR acceleration, a broken power-law spectrum 
with $\alpha_2\! \simeq \! 2.44$ above and $\alpha_1\! =\! 2.0$ 
below 5\,TeV gives 
the smallest possible $S\! \approx\!  4.9$.
A more plausible SNR spectrum with $\alpha_1\! =\! 2.2$ predicts 
$S\! \approx \! 24$.

Integration of the proton energy 
distribution formed by diffusion from a point `impulsive' 
source \citep{aav95} along the line of sight  gives
% \begin{equation}\label{distribution}
$ n(E,\theta,t) = N_0(E) \pi^{-1} l_{\rm dif}^{-2} 
\exp[-\zeta(E,\theta)/2 ]$,
% \end{equation}
where $l_{\rm dif}\! =\! \sqrt{2 D(E) t}$ is 
the mean diffusion length (along each of the axes) for a particle with 
energy $E$,  diffusion coefficient 
 $D(E)\! =\! D_{27} \,10^{27} (E/10\,\rm GeV)^{\delta}\,\rm cm^2\, s^{-1}$ 
in a power-law approximation, and 
$\zeta(E,\theta)\! \equiv \! (\theta \, d /l_{\rm dif})^2$.
Integrating over a circle of angular radius $\theta_{s}$, we 
derive a proton energy spectrum $ N(E;\theta \leq \theta_s) = N_0(E) \, 
[1 - \exp(-\zeta(E,\theta_s)/2) ]$. 

The observed spectral index $\alpha_\gamma \! =\!  2.44$ 
is significantly steeper than $\alpha_1\! \approx \! 2.1$-2.2 
expected for a typical SNR. 
We explain this steepening as an unavoidable result of energy 
dependent
diffusion. Given the HESS point spread function (PSF), 
the derived $N(E;\, \theta\! \leq \! \theta_s) $ suggests
$\zeta_0 \! \equiv \!  \zeta(10\,\rm TeV, 9^{\prime}.6) \! \sim \! 1$.
Calculations nicely fit the data
for $\zeta_0 = 0.61$, and exclude $\zeta_0>0.8$ or $\zeta_0 < 0.45$
(see Fig.\,1).
This defines the following key-relation
between the source distance and its age $t = t_{\rm kyr} \, 10^3\,\rm yr$
%\footnote
{ 
(we count $t$ from the time when the first light 
from the source reached us, and neglect the source expansion during the 
light crossing time $2 l_{\rm dif}/c \ll t$)\,}: 
\begin{equation}\label{distance}
d_{\rm kpc}=0.935\times 10^{3\delta/2}(D_{27}\, t_{\rm kyr})^{1/2} 
\, \zeta_{0}^{1/2} \, .
\end{equation}
This constrains the energy required to explain both the spectral and 
the spatial appearance of the source: 
\begin{equation}\label{angenergy}
U_{p}\approx 1.1 \times 10^{49+3\delta} (1+S)\, D_{27}\, t_{\rm kyr}\, 
n_{\rm H}^{-1} \zeta_0\,\, \rm erg.
\end{equation}
The diffusion coefficient in the ISM should have 
$\delta \! \simeq 0.5$ to explain the CR spectral index
$\alpha_{CR}\! = \! \alpha_{1} \! +\! \delta \approx \! 2.7$.
A typical value of $D(E)$ at $10\,\rm GeV$ is $D_{27}\gtrsim 10$ 
\citep{ber90,ps98}. A lower limit of $D_{27}\gtrsim 0.3$ can be derived from 
the grammage $X\equiv m_p n_{\rm H}c t_0\simeq 10\,\rm g\, cm^{-2}$ 
that explains the 
observed composition of CR nuclei \citep{buck94}, and does not allow CRs
to stay for more than $t_0\simeq 10^7\,\rm yr $ in the Galactic disk at 
heights $h_{\rm disk}\lesssim 150\,\rm pc$ where 
$n_{\rm H}\gtrsim 0.5\,\rm cm^{-3}$ \citep{ps98}.
Values of $D_{27}$ much smaller than the typical ones for the ISM 
would imply an enhanced level of turbulence and can be excluded, 
as the associated enhanced magnetic field would result in
radio emission exceeding the flux upper limit derived below.

 \begin{figure}[t]
 \epsscale{0.95}
 \plotone{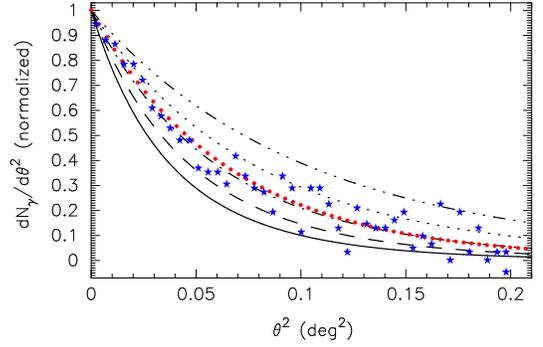}
 \caption{
 Observed (stars) and modelled (lines) 
 angular distributions of $\gamma$-rays from HESS J1301-631 (normalized 
 at $\theta=0$). From bottom to top, the lines show the distributions
 for $\zeta_0= 1.22$, 0.862, 0.61, 0.43, and 0.305, 
 convolved with the PSF   
 $\propto a \exp[-0.5(\theta/ 2\sigma_{1})^2] + b
  \exp[-0.5(\theta/2\sigma_{1})^2]$,
 with parameters  $\sigma_1$, $\sigma_2$ and $a/b$ derived from
 the PSF plot in \citet{hess1303}. 
 The full dots show the best-fit to the experimental data in the form of   
 PSF-convolved 2D Gaussian distribution with width  
 $\theta_{\rm src} = 9.6^\prime$.
 }
 \label{f1}
 \end{figure}

With these results, we are now in the position to argue that 
the source is a GRBR and not a typical SNR.
Using $D_{27}\gtrsim 0.3$, we constrain the source age   
to $t_{\rm kyr}\lesssim 0.2\, d_{\rm kpc}^2$. 
%%%
   Even for a SNR with $\alpha_1\!=\! 2.0$, an 
age $t_{\rm kyr} \! \lesssim \! 1.8$, and location at $d\! \sim \! 3 \,\rm kpc$ 
in the ISM with $n_{\rm H} \! \sim \! 1\,\rm cm^{-3}$,  
equation (1) predicts 
$U_{p}\! \geq \! 6.4 \!  \times \! 10^{50} n_{\rm H}^{-1}\, \rm erg$. 
For a typical 10\% CR injection efficiency by a SNR, this requires 
a kinetic energy $U_{\rm kin}\! \gtrsim \! 6\! \times \! 10^{51}\,\rm erg$. 
In principle, 
such energies cannot be excluded for some extremely   
powerful supernovae (SNe) of type Ib/c or IIn (which 
in fact might be GRBs)
that constitute $\lesssim \! 1\%$ of all SNe \citep{svesh03}.
This implies a total of 
$ N_{\rm SN} \lesssim  (0.1$-$0.3) t_{\rm kyr}$ such events
in our Galaxy. The expected number of such SNRs at a distance 
$d_{\rm kpc} \approx 3$ is only $\xi \! \simeq  N_{\rm SN} 
(d_{\rm kpc}/15)^2 \lesssim (0.7$-$2.1)\! \times 10^{-2}$.
Furthermore, the non-detection of the shell  of such powerful, close 
and young SNR in the radio
and X-ray bands is difficult to explain.

An assumption that the source is an SNR inside a molecular cloud with 
$n_{\rm H} = 56\,\rm cm^{-3}$ at $d=2.1\,\rm kpc$ \citep{hess1303} 
 relaxes the extremely high SNR energy requirement 
and increases $\xi $ to acceptable values.
However, the problem of the missing shell of this
still powerful and young SNR in such a dense medium persists.  
Even for a total energy in relativistic electrons of  
about $1\%$ of the protons, which is a typical lower limit for
an SNR, the synchrotron flux would significantly exceed  
the flux upper limit at 5 GHz (see Fig.~3).

\begin{figure}
 \epsscale{1.}
 \plotone{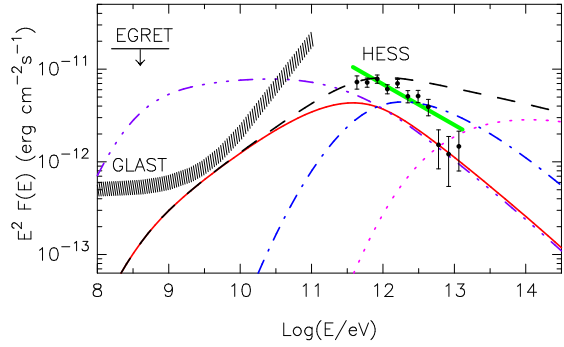}
 \caption{
 Hadronic ($\pi^0$-decay) model of the $\gamma$-ray 
 emission from HESS J1301-631, calculated for a GRBR of age 
 $t=1.5\times 10^4\,\rm yr$ at $d=12\,\rm kpc$ in the ISM with
 $n_{\rm H}=1\,\rm cm^{-3}$. Calculations assume a broken, ``GRBR type'',  
 power-law proton spectrum, with $\alpha_1=1.5 $ below 
 $E_{\rm brk}=5\,\rm TeV$. 
 The thick dashed line shows the predicted energy flux  
 when {\it all} emission within one degree from the source center is 
 considered. 
 The spectrum is considerably harder than the measured one 
 (data points and the bar).
 Owing to technical difficulties associated with the determination
 of the energy spectrum of an extended source, the H.E.S.S.\ collaboration 
 derived the spectrum using only photons detected within the central 
 $13.4^\prime$ radius and scaled it according to the total observed flux.
 We predict that the energy spectrum derived using all the TeV photons 
  will follow the dashed line. 
 The solid line shows the flux from the intrinsic 
 $\theta \leq 10^\prime$ region (this effectively translates to $13.4^\prime$ 
 size after correcting for the PSF). 
 The dashed-dotted line and the dotted lines show the 
 fluxes from 10$^\prime$-25$^\prime$ and 
 a 25$^\prime$-1$^\circ$ annuli, respectively.
 The 3-dot-dashed curve is the flux from the inner 
 $10^\prime$ angular region for an SNR-type spectrum of CRs,  
 $\alpha_1 =2.0$.
 Upper flux limit of EGRET for HESS J1303-631 \citep{mh05}
 and the 1-yr flux sensitivity of GLAST  are also shown. 
 }
 \label{f2}
 \end{figure}

For a source age of $t\! \gtrsim \! 10^4\,\rm yr$, equation 
(\ref{distance}) moves the source to $d\gtrsim \! 10\,$kpc and 
increases dramatically the energy requirements for an 
SNR that accelerates particles by non-relativistic shocks. 
The full radius of the source
(detected up to $\theta\lesssim\! 25^\prime$) 
increases to $\gtrsim \! 70 \,\rm pc$ implying that it is 
located either in ordinary ISM 
% with $n_{\rm H}\! \sim \! 1\,\rm cm^{-3}$ 
or {\it entirely} inside (to have a quasi-{\it uniform} profile for $n_{\rm H}$)
a giant molecular cloud. 
Figure 2 shows the fluxes calculated for 
$n_{\rm H}\! =\! 1\,\rm cm^{-3}$, $d\! =\! 12\,\rm kpc$ and 
$t\! =\! 1.5\!\times\! 10^4\,$yr. 
Two different proton energy spectra are considered. 
One implies an `extreme SNR', with  $\alpha_2\! =\! 2.2$ at 
$E \!\geq \! E_{\rm brk} =\! 5\,\rm TeV$ but 
$ \alpha_1\! =\! 2.0$ at 
$E\! < \! E_{\rm brk}$. The total energy derived is 
$U_{p.\rm SNR}\!=$ $ 8.1\! \times \! 10^{51}n_{\rm H}^{-1} \rm \, erg$.
The `GRBR' spectrum assumes $\alpha_1\! = \! 1.5$  implying protons 
accelerated  by a relativistic shock.
For qualitative estimates the break energy could be defined as 
$E_{\rm brk}\! \sim \! 0.5 \Gamma_{1}^2 \,$GeV, where $\Gamma_1$ is the shock 
Lorentz-factor at the time when most of the initial shock 
energy is transferred to relativistic protons (see Sec.3).
Fitting the data we find $\Gamma_{1}\! \lesssim \! 100$.
Thus $\Gamma_1$ can be as high as the {\it initial}  
$\Gamma_0\! \gtrsim\! 100$  
of GRB jets \citep{mesz02}.  Figure 2 assumes $\Gamma_1=100$. 
This reduces the total energy required for the GRBR model to 
$U_{p.\rm GRBR}\! =\! 4.1\! \times \! 10^{51} n_{\rm H}^{-1}\,\rm erg$.
%\footnote{ 
Values of $E_{\rm brk} \! < \! 5\,\rm TeV$, which are not excluded, 
would lead to an increase  
$U_{p.\rm GRBR}\propto \! (5\,{\rm TeV}/E_{\rm brk})^{0.2}$ (for 
$E_{\rm brk}\! \gtrsim \! 0.1 \, \rm TeV$), and even slower 
for $U_{p.\rm SNR}$.
%}

The HESS collaboration discussed the giant ($r\simeq 144\,\rm pc$) 
molecular cloud G303.9-0.4 \citep{grab88} with 
$n_{\rm H} \! \simeq \! 10$-$20\,\rm cm^{-3}$ located at 12 kpc from us 
as a possible host of the source. 
For the SNR-type spectrum, the minimum CR energy 
$U_{p.\rm SNR} =\,$(4-8)$\times \! 10^{50}\,\rm erg$
is problematic even for a very powerful SNR.
Furthermore, sub-relativistic protons alone would contain sufficient
energy for heating (at a rate 
$\sim \! 3\! \times\! 10^{35}\,\rm erg\, s^{-1}$),
and possibly also disintegration, of the cloud
via Coulomb losses.
Because of the ``low-energy cutoff'' expected below $E_{\rm brk}$, 
these constraints do not apply if HESS J1303-631 is a GRBR.
%%%    
It is also possible that the cloud is merely 
a line of sight coincidence.  
Even at a distance of $d =$15\,kpc a GRBR would still be able to 
power the inferred proton energy density in the ISM.

 \begin{figure}
 \epsscale{1.}
 \plotone{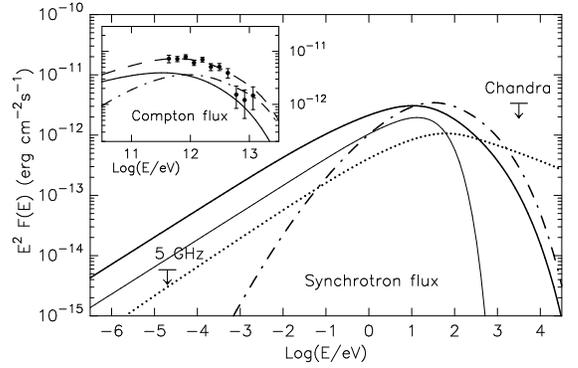}
 \caption{ 
 Synchrotron fluxes (shown in the main frame) expected from HESS J1301-631
 in the case of a leptonic (IC) origin of TeV fluxes (shown in the inset).
 The fluxes expected from angular distances 
 $\theta \leq 10^\prime$ and $\theta > 10^\prime$ 
 are plotted with heavy solid and dot-dashed lines, respectively.
 The total IC flux (dashed line) is normalized to the 
 observed flux at 1\,TeV. The source age $t=10^4\,\rm yr$ and 
 $ B=3\,\rm \mu G$  are assumed. 
 The thin solid curve shows the synchrotron flux from accelerated 
 (primary) electrons with total energy $1\%$ of relativistic 
 protons, to be expected from a SNR  at $d=2.1\,kpc$ in dense cloud with
 $n_{\rm H}=56\,\rm cm^{-3}$ \citep{hess1303}  and enhanced 
 level of turbulence and magnetic field, $B= 10\,\rm \mu \rm G$,
 where the observed TeV flux would be explained by the protons.
 The synchrotron radiation from secondary electrons is shown 
 by the full-dotted line.  
 }
 \label{f3}
 \end{figure}

Leptonic models require less power than hadronic models.
IC radiation of multi-TeV electrons on the CMB and Galactic 
far-infrared photons could reproduce the observed spectrum.
% as the Klein-Nishina effect on FIR photons modifies the 
% spectrum at $E_{\gamma} \gtrsim\! 1\,\rm TeV$. 
A broken power-law electron spectrum with $\alpha_{e.1} \! = \! 2.1$ 
below 10 TeV, and $\alpha_{e.2}\! =\! 3.2 $ above leads to 
the minimum required energy  
$U_{e}\! \simeq \! 2.4\! \times\! 10^{48}d_{\rm kpc}^2 \,\rm erg$. 
However,  the synchrotron radio flux of the leptonic model exceeds the 
flux upper limit that we derived from the Parkes-MIT-NRAO southern  
4850\,MHz survey \citep{cond93} even for an ISM magnetic field 
$B_{\rm ISM}\! = \! 3\,\rm \mu G$ (see Figure 3).
We estimate the limit by measuring a mean flux of  0.0032 Jy/beam 
in the $9^{\prime}.6$ radius region centered on the source. 
Calculating the flux in background regions of the same aperture, 
we estimate a standard deviation of 0.0033 Jy/beam. 
Using a beam width of $7^\prime$ FWHM, we then obtain a 5$\sigma$ upper limit 
 of 0.12\,Jy on the flux from within $9^{\prime}.6$ from the source centroid. 

We have also computed the IC fluxes  
from the secondary $\pi^\pm$-decay 
electrons and from $\beta$-decay electrons \citep{ikm04},
accounting also for their radiative energy losses. 
Both fluxes are almost two orders of magnitude lower than 
the $\pi_0$-decay flux. 
The ratio of synchrotron to IC fluxes depends 
only on the ratio of the magnetic field to the target photon energy
densities. Hence the radio flux from the secondary electrons for the 
GRBR model in Figure 2 does not violate the derived upper limit  
in so far as their IC emission is well below the observed TeV flux.   
% And the primary electrons in GRBRs are suppressed, 
% as we argue below. 
% The non-observation of synchrotron radiation 
% or a shell-type morphology while incompatible with the SNR hypothesis, 
% can be explained in the framework of the GRBR hypothesis.
%

\section{Discussion and predictions}
% Here we argue that 
A GRBR should differ from a SNR not only by the energetics, 
but also by three characteristics of its relativistic particle population.
First, the prompt differential spectrum of protons (and electrons) 
behind the relativistic shock cannot be a single power-law 
$N(E)\propto E^{-\alpha}$ with $\alpha \geq 2.0$ 
as in the case of nonrelativistic shocks.
As shocks with Lorentz-factor $\Gamma(t) \! \gg 1 \! $ boost
{\it all} particles they encounter to relativistic energies, 
the prompt spectrum should break to an index 
$\alpha_{1, \rm prompt} \! < \! 1$ below some energy 
$E_{\rm brk}^\prime \! \lesssim \! m_p c^2 \Gamma^2(t)/2$ 
(observer frame) to satisfy the laws of energy and 
particle number conservation across the shock \citep{katz94}. 
The final power-law index $\alpha_1 $ (presumably $ \!\lesssim \! 1.5 $) 
and the break energy $E_{\rm brk}$ are formed while the shock decelerates, 
and will depend on a number of processes,
such as diffusive escape of relativistic protons from the shell,   
their adiabatic cooling and reacceleration, 
etc. 
 At $E \! > \! E_{\rm brk}$ a spectral index $\alpha_2 \! \approx \! 2.2$ 
is expected to form in Fermi-type processes \citep{kirk99}.
For the current data, $E_{\rm brk}$ could be as high as 5\,TeV 
(in Figure 2), but it could also be smaller,
leading to GeV $\gamma$-ray fluxes detectable by GLAST.

The second feature of GRBRs that we predict is that a substantial fraction of 
the initial shock energy is eventually transferred to high-energy 
protons, while relativistic electrons would contain only a tiny
fraction of it at times of non-relativistic evolution of the GRBR shell.
A GRB shock becomes non-relativistic, and also quasi-spherical, 
on timescales $\lesssim 1\,\rm yr$ at distances $\lesssim 0.1\,\rm pc$ 
\citep{lw00}. At  this stage,
magnetic fields of $\sim \! 0.1\,\rm G$ \citep{bkf04} 
are still high and will (synchrotron) cool electrons to a few GeV  
within $\sim$1\,yr. As the initial kinetic energy  
(5-10)$\times\! 10^{51}\,\rm erg$ of the jets is much larger 
than the radiated (and the magnetic field's) energy, 
by then most of it should be in relativistic protons.
The evolution of a non-relativistic shell that is dominated by the 
relativistic energy, and not by the kinetic energy of the shock, 
requires a fully relativistic treatment and remains unexplored. 
Diffusion of relativistic protons (thus escaping adiabatic losses) 
upstream may broaden the shock precursor. Because of the preferential 
acceleration of particles with Lorentz factors larger than the 
precursor \citep{be99}, energetic protons will readily be accelerated 
while acceleration of electrons will be suppressed.
This would also apply to the spherical ``supernova''-type component 
(which, however, may not be developed by GRBs at all, see \citet{woosley})
as this non-relativistic shock has to pass through the central 
$\lesssim 10 \, \rm pc$ region dominated by the energy of relativistic 
protons from the (quasi-spherically opened up) GRB jets.
We also note that in the presence of relativistic protons, 
nonrelativistic shocks may evolve (and dissipate) much faster than 
predicted by the standard Sedov-Taylor solution  for SNR shocks propagating 
in ``cold'' plasmas \citep{topt00}. 
% The theories of deceleration of GRB jets and particle acceleration in 
% relativistic shocks are still plagued by serious uncertainties. We hope 
% that this paper stimulates further theoretical research of these problems.

Third, considerable energy, $\sim\!(10$-$20)\,\%$, of the accelerated
protons is channeled into collimated beams of neutrons 
\citep{ad03,da03} from    
interactions $p\! +\! \gamma \! \rightarrow \!  n \! +\! \pi^+$ 
with X-ray photons from the GRB. 
First, this mechanism assures a minimum level of total energy of 
the protons that avoid adiabatic losses in the shell, 
$U_{n}\gtrsim 10^{51}\,\rm erg$. Furthermore, neutrons with 
energies $E_n$ will $\beta$-decay into protons on 
length scales of $l_{\rm n}(E_n) \simeq (E_n/100\,\rm TeV) \,\rm pc$.
A characteristic signature of a GRBR might then be an 
elongation of the TeV $\gamma$-ray emission due to conversion 
of the initial double-sided neutron beam into protons, 
which will subsequently be injected into the diffusion
process along the jet path on length scales $\sim \!1$-$10\,\rm pc$ 
(for $E_n\lesssim 1\,\rm PeV$) .
Depending on the contribution of neutrons to the high-energy proton 
population, the two-dimensional Gaussian profile may have an aspect 
ratio of up to
$A=(\sigma_\parallel-\sigma_\perp)/(\sigma_\parallel+\sigma_\perp)$ 
with $\sigma_\perp= l_{dif}(E,t)$ 
and $\sigma_\parallel = \sqrt{l_{\rm n}^2(E)+l_{\rm dif}^2(E,t)}$.
For a source size $ l_{\rm dif}\gtrsim 10\,\rm pc$ this would 
predict $A(E)\! \sim \! (0.01$-0.1). 
However, neutron beams can also affect the shape of the 
GRBR nebula indirectly via driving (after  $\beta$-decay) the ISM, and 
stretching the ambient magnetic field,
along the jet direction \citep{ad03}. 
A faster diffusion of the bulk of protons in this direction 
could result in $A(E)\gg 0.1$. 
Remarkably, in both cases the aspect ratio should increase at higher energies.

The HESS collaboration determined the source spectrum with  
the photons from the bright central part of the source 
(within 13.4$^\prime$ from the 
centre). While our model explains this spectrum, it
predicts that the energy spectrum will be significantly harder 
when determined for the entire source (see Figure 2).  
Another specific signature of a GRBR is the unusually hard energy 
spectrum at $E\leq E_{\gamma,\rm brk}\! \sim \! 10$-100\,GeV.
GLAST will be able to verify or falsify this prediction. 
If HESS~J1303-631 is a SNR, GLAST will easily detect the source 
at the flux level close to the upper limit of EGRET \citep{mh05}. 
However, a weak detection (confirming the hard spectrum)
or non-detection by GLAST would signify the tell-tale detection of the 
signatures of proton acceleration by relativistic shocks of a GRB.

Confirmation of HESS J1303-631 and at least one other extended unidentified 
TeV source, such as TEV J2032+4130 \citep{hegra02,whipple}, as GRBRs
would confirm the high rate, of order $10^{-4}\,\rm yr^{-1}$, of Galactic 
GRBs, and would make it more likely that GRBs had a direct impact on life on 
Earth \citep{mel04}.

\vspace{1mm}
We thank Dr. C. Dermer, Dr. P. M\'esz\'aros, and the referee 
for helpful comments. 
 AA acknowledges the hospitality of the Physics Department and of the 
McDonnell Center for Space Sciences of the Washington University.

 \end{document}